\documentclass[preprint,showpacs,preprintnumbers,amsmath,amssymb]{revtex4}

\usepackage{graphicx}
\usepackage{dcolumn}
\usepackage{bm}

\makeatletter

\newcommand{\Rmnum}[1]{\expandafter\@slowromancap\romannumeral #1@}
\makeatother

\begin{document}


\title{A Realization of Effective SUSY with Strong Unification}

\author{Chun Liu and Zhen-hua Zhao}

\email{liuc@itp.ac.cn, zhzhao@itp.ac.cn}

\affiliation
{Institute of Theoretical Physics, Chinese Academy of Sciences, \\
and State Key Laboratory of Theoretical Physics,\\
P. O. Box 2735, Beijing 100190, China}

\date{\today}

\begin{abstract}
A natural model of realizing the effective supersymmetry is presented.
Two sets of the Standard Model-like gauge group $G_1\times G_2$ are
introduced, where $G_i=SU(3)_i\times SU(2)_i\times U(1)_i$, which
break diagonally to the Standard Model gauge group at the energy scale
$M \sim 10^7$ GeV.  Gauge couplings in $G_1$ are assumed much larger
than that in $G_2$.  Gauge mediated supersymmetry breaking is adopted.
The first two generations (third one) are charged only under $G_1$
($G_2$).  The effective supersymmetry spectrum is obtained.  How to
reproduce realistic Yukawa couplings is studied.  Fine-tuning for an
126 GeV Higgs is much reduced by the large $A$ term due to direct
Higgs-messenger interaction.  Finally, $G_2$ is found to be a
non-trivial realization of the strong unification scenario in which
case we can predict $\alpha_s(M_Z)$ without real unification
\end{abstract}


\maketitle

\section{Introduction}

A Standard Model (SM)-like Higgs particle of 126 GeV has been discovered
at the LHC \cite{Higgs search}.  If we are insisting on naturalness of
the SM, this discovery strengthens motivation for the low energy
supersymmetry (SUSY) which stabilizes the Higgs mass at the electroweak
(EW) scale.  However, there is yet no definite sign of sparticles after
the integrated luminosity has reached 20 fb$^{-1}$ at $\sqrt s = 8$
TeV.  This null result for the SUSY search sets a lower bound for the
first two generation squarks -- $m_{\tilde Q_{1,2}} > 1$ TeV, while
stops/sbottoms with a mass about 500 GeV are still allowed
\cite{supersymmetry search}.  Noticing that naturalness sets an upper
bound for sparticle masses as 1 TeV.

SUSY, if it is relevant to the EW physics, should be beyond its simplest
version.  Actually it was noted long time ago that naturalness only
requires the third generation sfermions and particles (gauginos and
Higgsinos) that interact significantly with the Higgs to have sub-TeV
masses, while the first two generation sparticles can be heavy up to 20
TeV \cite{esusy1,esusy2,effective susy}.  Such sparticle spectra also
alleviate the SUSY FCNC problem.  This phenomenological scenario is
dubbed Effective SUSY by Cohen {\it et al.} \cite{effective susy}.
Nowadays this Effective SUSY has become one of the main scenarios to
reconcile naturalness with the null SUSY search \cite{natural susy}.

In this paper, we realize Effective SUSY through modifying models of
Refs.\cite{kl,liu}.  In those models we have introduced two sets of
SM-like gauge groups $G_1 \times G_2$ where
$G_i=SU(3)_i \times SU(2)_i \times U(1)_i$.  At the TeV scale, $G_1$
is strongly and $G_2$ is weakly interacting, respectively.  They break
diagonally to the SM gauge group.  SUSY breaking is due to
$G_1 \times G_2$ gauge mediation.  Furthermore, $G_2$ is of strong
unification \cite{liu}, namely its gauge coupling constants have a
common Landau pole at the unification scale
\cite{strong unification1,strong unification2,strong unification3}.
In that model, all the three generations were put in $G_2$, that
did not result in any sparticle splitting.  To make sparticle
splitting, the first two generations have to be treated differently
from the third one.  In this work, the first two generations are put in
$G_1$ and the third in $G_2$.

Among other things, we need to solve the following problems.  First,
to generate Yukawa interactions between the first two generations and
the third generation.  This is because three generations are in
different gauge sectors from the beginning.  Then, to reduce fine
tuning of the 126 GeV Higgs.  In conventional gauge mediated SUSY
breaking (GMSB) \cite{gmsb}, fine-tuning seemed unavoidable for a 126
GeV Higgs.  Furthermore, to re-examine strong unification.  This is
needed due to all the changes in the particle content, and conditions
of strong unification are quite subtle.

We notice that a similar model was proposed by Craig {\it et al.} \cite{split families unified1,split families unified} in which the first two
generations were also put in $G_1$ and the third one in $G_2$.
However, there are several main differences.  First is about $G_1$.  In
the model of \cite{split families unified}, $G_1$ gauge interactions
are weakly coupled at the TeV scale, whereas our $G_1$ is superstrong.
Second is for $G_2$.  Their $G_2$ gauge interactions unify weakly in
the sense of ordinary grand unification, our $G_2$ is of strong
unification
\cite{strong unification1,strong unification2,strong unification3}.
Third is about GMSB.  They only use a messenger for $G_1$, and we have
messengers both for $G_1$ and $G_2$.  And finally we need to use direct
Higgs-messenger interaction to reduce fine-tuning of a 126 GeV Higgs.
These differences make this model qualitatively different. \par

There were a few other ways to realize Effective SUSY
\cite{esusy model,radiatively generated}.  Some of them used an extra $U(1)$
gauge group which contributes larger masses to the first two generation
sparticles than to the third generation ones.  Usually, this $U(1)$
symmetry suppresses Yukawa couplings for the first two generations
compared to that for the third generation, giving an explanation for
the fermion mass hierarchy.  Some other works assume particular
boundary conditions at a high scale, then employ the technique of
renormalization group \cite{radiatively generated}.

The paper is organized as follows.  Our model will be given in the
next section, where flavor physics for
fermions and the problem of
naturalness in light of a 126 GeV Higgs will be discussed.
After all particle contents
and the mass spectrum have been fixed, the prediction for
$\alpha_s(M_Z)$ will be calculated by means of strong unification in Section \Rmnum 3.
The final section summarizes our results and gives discussions.

\section{The Model}

We consider a SUSY model with two sets of the SM-like gauge group $G_1$
and $G_2$ where $G_i=SU(3)_i\times SU(2)_i \times U(1)_i$.  The first
two and the third generation of matter transform under $G_1$ and $G_2$,
respectively.  The two Higgs doublets $H_u$ and $H_d$ are in $G_2$.
The other fields include SUSY breaking messengers and the Higgs fields
which break $G_1\times G_2$ into the SM.  For convenience, we will use
field representations under $SU(5)$ to illustrate their representations
under $SU(3)\times SU(2)\times U(1)$.

The GMSB mechanism is employed.  Two sets of messenger fields $T_1$
$(\bar T_1)$ and $T_2$ $(\bar T_2)$ are introduced.  They transform
nontrivially under $G_1$ and $G_2$, respectively.  Without losing
generality, we will focus on the quark/squark sector.  At the scale of
SUSY breaking, squarks have following masses,
\begin{equation}
\label{1}
m^2_{\tilde Q_{1,2}}\sim (\frac{g_1^2}{16 \pi^2}\frac{F}{M})^2\,,
\hspace{0.5cm} {\rm and} \hspace{0.5cm}
m^2_{\tilde Q_3}\simeq (\frac{g_2^2}{16 \pi^2}\frac{F}{M})^2\,.
\end{equation}
$M$ stands for the messenger scale and $\sqrt F$ is the measure of SUSY
breaking.  $g_1$ and $g_2$ represent coupling constants for $G_1$ and
$G_2$ respectively. Therefore, we can realize
Effective SUSY sparticle spectrum by requiring $g_1$ to be much larger
than $g_2$.  Note that Eq.(\ref{1}) for $m^2_{\tilde Q_{1,2}}$ is not
exact, because $g_1$ is too large.

A pair of Higgs fields $\Phi$ and $\bar\Phi$ charged under
$G_1\times G_2$ as $\textbf{5}\times\bar{\textbf{5}}$ and
$\bar{\textbf{5}}\times\textbf{5}$ is introduced.
$\Phi$ and $\bar\Phi$ have a mass $M_\Phi$, and they obtain vacuum
expectation values (VEVs) as
$\langle\Phi\rangle=\langle\bar\Phi\rangle=V I_2\times I_3$, where
$V\sim M_\Phi$, $I_2$ and $I_3$ are the unit matrix in the subspace of
$SU(2)_1\times SU(2)_2$ and $SU(3)_1\times SU(3)_2$, respectively.
As a result, $G_1\times G_2$ break diagonally to the SM gauge group
$SU(3)_c\times SU(2)_L\times U(1)_Y$.  Below the scale of $M_\Phi$, the
effective theory of this model looks like Effective SUSY with following
relations among gauge coupling constants,
\begin{equation}
\frac{1}{g_s^2}=\frac{1}{g_{s1}^2}+\frac{1}{g_{s2}^2}\,,\hspace{0.5cm}
\frac{1}{g^2}=\frac{1}{g_{1}^2}+\frac{1}{g_{2}^2}\,,\hspace{0.5cm}
\frac{1}{g^{\prime 2}}=\frac{1}{g^{\prime 2}_1}+\frac{1}{g^{\prime 2}_2}
\,.
\end{equation}
Because of the relation $g_1^2\gg g_2^2$, SM gauge couplings are almost
fully determined by that in $G_2$.  More details about breaking of
$G_1\times G_2$ can be found in Ref. \cite{liu}.  While this model in
many aspects is similar to that of Ref. \cite{liu}, new features come in
because of separation of three generations.

Before $G_1\times G_2$ breaking, three generations are put in different
gauge sectors, so there is no marginal operators giving Yukawa couplings
between the first two generation matter and the third generation ones.
However, higher-dimensional operators such as $H_d Q_3 \Phi_{d\bar d} \bar d_{1(2)}$ are allowed by $G_1\times G_2$ symmetry.
Here $\Phi_{d\bar d}$ is the component of the Higgs $\Phi$
with quantum numbers
$(3,1,-\displaystyle\frac{1}{3})*(\bar 3,1,\displaystyle\frac{1}{3})$
under
$SU(3)_1\times SU(2)_1\times U(1)_1\times SU(3)_2\times SU(2)_2\times U(1)_2$.
This kind of operators can be produced by integrating out appropriate heavy fields, just like the Froggatt-Nielsen mechanism \cite{FN}.
As a result, they are suppressed by $M$ (mass scale for the heavy fields that have been integrated out) $\displaystyle \frac{1}{M}H_d Q_3 \Phi_{d\bar d} \bar d_{1(2)}$. When $G_1\times G_2$ is spontaneously broken i.e. $\Phi_{d\bar d}$ gets VEV, there will be terms like $\displaystyle \frac{V}{M}H_d Q_3 \bar d_{1(2)}$ that lead to Yukawa interactions between 1st/2nd generation fermions and 3rd generation fermions.
Taking $\displaystyle \frac{V}{M}$ a small quantity $\sim 0.1-0.01$, the mass hierarchy between the third generation and the first two is obvious.  Roughly
speaking, this paves the way to obtain the realistic fermion mass
pattern, mixing and CP violation.
We can say hierarchy among three generation fermions and that among three generation sfermions are closely connected to each other in this model.

This approach has been discussed in Ref. \cite{split families unified}.
There, a vector-like $\textbf{5}$  representation is introduced as the mediator that is integrated out.
Similarly, in $G_2$, we introduce a full vector-like generation $(L,\bar d)$ and
$(Q,\bar u, \bar e)$ as representation $\bar{\textbf{5}}$ and
$\textbf{10}$, respectively.
Masses of these vectorlike fields are
taken to be of the same order $M_\Psi$.

In MSSM, the SM-like Higgs has the following mass after including one-loop radiative corrections induced by stops,
\begin{equation}
m_h^2=m_Z^2 \cos^2{2\beta}+\displaystyle \frac{3 m_t^4}{4 \pi^2 v^2}[\log{\displaystyle \frac{m_{\tilde t}^2}{m_t^2}}+\displaystyle \frac{X_t^2}{m_{\tilde t}^2}(1-\displaystyle \frac{X_t^2}{12 m_{\tilde t}^2})],
\end{equation}
where $X_t=A_t-\mu \cot \beta$. To obtain a 126 GeV Higgs, we either need multi-TeV stops which lead to
severe fine-tuning, or turn to the large $A_t$ scenario
in which case sub-TeV stops are enough.  In conventional GMSB,
contribution of $A_t$ is negligible and fine-tuning induced by too
heavy stops seems to be unavoidable.
To preserve naturalness, we will produce a large $A_t$ term by extending conventional GMSB.

We choose the messenger $T_2(\bar{T}_2)$ to be the representation
$\textbf{10}$ ($\bar{\textbf {10}}$).  It is found that
\cite{li yang}, a large $A_t$
term can be produced without a large Higgs mass.  This is due to direct
interaction between the Higgs and the messenger in the superpotential,
\begin{equation}
y H_u T_2^Q T_2^{\bar u} \,,
\end{equation}
where $T_2^Q$ and $T_2^{\bar u}$ are components of $T_2$ that have same
gauge quantum numbers as $Q_3$ and $\bar u_3$, respectively.  It should
be pointed out that this term is the only one of direct interaction
between messengers and ordinary matters by employing the messenger
parity.  One-loop contribution to the soft term
$H_u\tilde Q_3\tilde{\bar u}_3$ is extracted from wave function
renormalization for the superfield $H_u$
\cite{wave function renormalization},
\begin{equation}
A_t \sim - \displaystyle \frac{y^2 y_t}{16 \pi^2} \frac{F}{M}\,.
\end{equation}
In Ref. \cite{eiy}, the same way was used to produce a large $A_t$ term
except that there it was $Q_3$ and $\bar u_3$ instead of $H_u$ that
interact directly with the messenger.
More issues about this way of producing large $A_t$ term can be found in \cite{li yang}\cite{es}.

\section{Strong Unification}

We will consider unification of gauge coupling constants.  Because the
SM gauge couplings are almost fully determined by those of $G_2$, what
we will really care about, above $G_1*G_2$ breaking scale, is the $G_2$
gauge coupling constants.  So far
quite a few new fields have been introduced.  The particle content and
the mass spectrum are summarized in the following with emphasis on that
which are charged under $G_2$.  There is only one chiral generation in
$G_2$.  The two Higgs doublets are in $G_2$.  The bi-fundamental Higgs
$\Phi$ ($\bar\Phi$) is charged under $SU(5)_1\times SU(5)_2$ as
$\textbf{5}\times\bar{\textbf{5}}$ ($\bar{\textbf{5}}\times\textbf{5}$)
with a mass $M_\Phi$.  The messengers $T_2$ $(\bar T_2)$ is charged
under $G_2$ as representation $\textbf{10}$ $(\bar{\textbf {10}})$ with
a mass $M$.  Besides, there is an extra vector-like generation charged
under $G_2$ as representation $\bar{\textbf{5}}+\textbf{10}$
$(\textbf{5}+\bar{\textbf{10}}$) with a mass $M_\Psi$.  For simplicity
and definiteness, we will identify $M_\Phi$ and $M_\Psi$ with $M$ in the
following analysis.

Below $G_1*G_2$ breaking scale $M$, this model is just that of the minimal SUSY SM, so SM gauge couplings running can be calculated in the usual way.
At the scale $M$, SM gauge couplings are identified with that in $G_2$.
Above the scale $M$, since there are
so many complete representations of $SU(5)_2$, while the unification
energy scale does not change, gauge couplings in $G_2$ grow so fast that
they may come across their Landau poles as they evolve to the
unification scale.  The situation, where gauge couplings reach their
common Landau pole, is named as strong unification
\cite{strong unification1,strong unification2,liu}.
Using strong unification to predict gauge couplings at the
EW scale seems unreasonable because of the strong coupling domain where
the perturbative method is not reliable.
However, as shown in
Ref. \cite{strong unification2}, ratios of gauge couplings in $G_2$ will reach
their infra-fixed points at the scale $M$.  Thus, we can determine SM gauge
couplings at the EW scale, with ratios of gauge couplings in $G_2$ at the scale $M$ as boundary
condition where perturbative calculation already works.

Boundary conditions for gauge couplings of $G_2$ at the scale $M$ satisfy the following relation \cite{strong unification2},
\begin{equation}
\alpha_2^\prime(M) b_2^\prime=\alpha_{2}(M) b_{2}=\alpha_{s2}(M)b_{s2},
\end{equation}
where $\alpha_2^\prime(M)=\displaystyle\frac{g_2^{\prime 2}(M)}{4\pi}$
and so forth.  $b_2^\prime=\displaystyle\frac{73}{5}$, $b_{2}=9$ and
$b_{s2}=5$ are one-loop beta functions above the scale $M$ for
$g_2^\prime$, $g_{2}$ and $g_{s2}$, respectively.
In the following, we will first calculate the prediction for $\alpha_s(M_Z)$ in a simply way to illustrate the usage of strong unification, and then take into consideration two-loop contributions and low-scale threshold effects induced by sparticles' mass splitting among different generations.

In the first case, $\alpha^\prime(M)$ and $\alpha(M)$ can be determined by
$\alpha^\prime(M_Z)$ and $\alpha(M_Z)$ through the following
equations,
\begin{equation}
\alpha^{\prime-1}(M)=\alpha^{\prime-1}(M_Z)-\frac{b^\prime}{2\pi}\ln{\frac{M}{M_Z}}
\end{equation}
\begin{equation}
\alpha^{-1}(M)=\alpha^{-1}(M_Z)-\frac{b}{2\pi}
 \ln{\frac{M}{M_Z}}.
\end{equation}
With Eq.(6), we can get $M\sim 10^8$ GeV and $\alpha_s^{-1}(M)=15.17$.
Finally, $\alpha_s(M_Z)$ is calculated to be 0.119 as follows,
\begin{equation}
\alpha^{-1}_s(M_Z)=\alpha^{-1}_s(M)+\frac{b_s}{2\pi}\ln{\frac{M}{M_Z}}.
\end{equation}

After inclusion of low scale threshold effects and dominated two-loop contributions, Eqs.(7-9) will be replaced by the following equations,
\begin{equation}
\begin{aligned}
\alpha^{\prime-1}(M)=\alpha^{\prime-1}(M_Z)- \frac{\tilde{\tilde {b^\prime}}}{2\pi}
 \ln{\frac{m_{\tilde Q_3}}{M_Z}}-\frac{\tilde {b^\prime}}{2\pi}
 \ln{\frac{m_{\tilde Q_{1,2}}}{m_{\tilde Q_3}}}
 -\frac{b^\prime}{2\pi}
 \ln{\frac{M}{m_{\tilde Q_{1,2}}}}\\
 \hspace{1.1cm}-\frac{1}{4\pi}\frac{b_{11}}{ b^\prime}\ln{\frac{\alpha^{\prime}(M)}{\alpha^{\prime}(M_Z)}}-\frac{1}{4\pi}\frac{b_{12}}{ b}\ln{\frac{\alpha(M)}{\alpha(M_Z)}}-\frac{1}{4\pi}\frac{b_{13}}{b_s}\ln{\frac{\alpha_s(M)}{\alpha_s(M_Z)}},
\end{aligned}
\end{equation}
\begin{equation}
\begin{aligned}
\alpha^{-1}(M)=\alpha^{-1}(M_Z)- \frac{\tilde{\tilde {b}}}{2\pi}
 \ln{\frac{m_{\tilde Q_3}}{M_Z}}-\frac{\tilde {b}}{2\pi}
 \ln{\frac{m_{\tilde Q_{1,2}}}{m_{\tilde Q_3}}}
 -\frac{b}{2\pi}
 \ln{\frac{M}{m_{\tilde Q_{1,2}}}}\\
 -\frac{1}{4\pi}\frac{b_{21}}{ b^\prime}\ln{\frac{\alpha^{\prime}(M)}{\alpha^{\prime}(M_Z)}}-\frac{1}{4\pi}\frac{b_{22}}{ b}\ln{\frac{\alpha(M)}{\alpha(M_Z)}}-\frac{1}{4\pi}\frac{b_{23}}{b_s}\ln{\frac{\alpha_s(M)}{\alpha_s(M_Z)}},
\end{aligned}
\end{equation}
\begin{equation}
\begin{aligned}
\alpha_s^{-1}(M)=\alpha_s^{-1}(M_Z)- \frac{\tilde{\tilde {b_s}}}{2\pi}
 \ln{\frac{m_{\tilde Q_3}}{M_Z}}-\frac{\tilde {b_s}}{2\pi}
 \ln{\frac{m_{\tilde Q_{1,2}}}{m_{\tilde Q_3}}}
 -\frac{b_s}{2\pi}
 \ln{\frac{M}{m_{\tilde Q_{1,2}}}}\\
 -\frac{1}{4\pi}\frac{b_{31}}{ b^\prime}\ln{\frac{\alpha^{\prime}(M)}{\alpha^{\prime}(M_Z)}}-\frac{1}{4\pi}\frac{b_{32}}{ b}\ln{\frac{\alpha(M)}{\alpha(M_Z)}}-\frac{1}{4\pi}\frac{b_{33}}{b_s}\ln{\frac{\alpha_s(M)}{\alpha_s(M_Z)}},
\end{aligned}
\end{equation}
with
\begin{equation}
 \begin{array}{llllll}
 \vspace{0.4cm}
 \tilde{\tilde {b^\prime}}=\frac{41}{10}&\hspace{0.3cm}\tilde{b^\prime}=\frac{79}{15}&\hspace{0.3cm}b^\prime=\frac{33}{5}&\hspace{0.3cm}b_{11}=\frac{199}{25}&\hspace{0.3cm}b_{12}=\frac{27}{5}&\hspace{0.3cm}b_{13}=\frac{88}{5}\\\vspace{0.4cm}
 \tilde{\tilde{b}}=-\frac{19}{6}&\hspace{0.3cm}\tilde{b}=-\frac{1}{3}&\hspace{0.3cm}b=1&\hspace{0.3cm}b_{21}=\frac{9}{5}&\hspace{0.3cm}b_{22}=25&\hspace{0.3cm}b_{23}=24\\\vspace{0.4cm}
 \tilde{\tilde{b_s}}=-7&\hspace{0.3cm}\tilde{b_s}=-\frac{13}{3}&\hspace{0.3cm}b_s=-3&\hspace{0.3cm}b_{31}=\frac{11}{5}&\hspace{0.3cm}b_{32}=9&\hspace{0.3cm}b_{33}=14
 \end{array}
\end{equation}
It is found that $\alpha_s(M_Z)\sim 0.117$ and $M \sim 10^7$ GeV, when $m_{\tilde Q_3}$ and $m_{\tilde Q_{1,2}}$ take typical value 1 TeV and 10 TeV respectively.
This value is very close to world average value $0.1184\pm0.0007$ \cite{experimental value}.

In the above discussion, we have taken the limit $\displaystyle\frac{g_1^2}{g_2^2}$ goes to infinity so that SM gauge coupling can be identified with that in $G_2$ at the Higgsing scale.
There will be several percents uncertainty in this identification, if we take into consideration that $\displaystyle\frac{g_1^2}{g_2^2}$ is finite which is several tens.
This uncertainty will affect the prediction for $\alpha_s(M_Z)$ substantially.
For example, with a typical value 20 $\sim$ 40 for $\displaystyle\frac{g_1^2}{g_2^2}$, $\alpha_s(M_Z)$ will have an uncertainty about 0.005.
However, if three gauge couplings in $G_1$ sector has the same ratio with the counterparts in $G_1$ sector,
\begin{equation}
\displaystyle\frac{g^\prime_1}{g^\prime_2}=\displaystyle\frac{g_1}{g_2}=\displaystyle\frac{g_{s1}}{g_{s2}}
\end{equation}
this uncertainty will disappear.
This is because the boundary condition Eq.(6) just depends on the ratio of gauge coupling constants.

\section{Summary and Discussion}

In summary, we have presented a model of realizing the Effective SUSY.
Two sets of the SM-like gauge group
$G_1\times G_2=SU(3)_1\times SU(2)_1\times U(1)_1\times SU(3)_2\times SU(2)_2\times U(1)_2$
have been introduced which break diagonally to the SM gauge group at the
energy scale $M \sim 10^7$ GeV.  Gauge couplings in $G_1$ have been
assumed much larger than that in $G_2$.  GMSB has been adopted.  The
first two generations (third one) are charged only under $G_1$ ($G_2$).
The Effective SUSY spectrum has been obtained naturally.  Fine-tuning
for an 126 GeV Higgs is much reduced.  With all the fields necessary and
their masses fixed, $\alpha_s(M_Z)$ can be predicted in the scenario of
strong unification.

Compared to our previous works \cite{kl, liu}, in addition to the
Effective SUSY spectrum, following new features have arisen. \par
(1) An extra vector-like generation charged under $G_2$ has been
introduced as mediator, so as to reproduce realistic Yukawa couplings
between the first two generations and the third generation, i.e. the
suitable fermion mass hierarchy and the CKM mixing matrix.  \par
(2) Fine-tuning for an 126 GeV Higgs is much reduced by a large $A_t$
term produced by direct Higgs-messenger interaction, because the
messenger for $G_2$ has been specified to be a {\bf 10} representation
under the $SU(5)$, which is absent in conventional GMSB.

The following three main aspects clarify differences of
Ref. \cite{split families unified} from our model.  \par
(a) In Ref. \cite{split families unified}, gauge couplings in $G_1$ and
$G_2$ were comparable.  Only a messenger for $G_1$ was introduced, and
the third generation sparticles could feel SUSY breaking only after the
breaking of $G_1\times G_2$, so that $m_{\tilde Q_3}$ was suppressed by
an additional factor $\displaystyle \frac{V}{M}$ in comparison with
$m_{\tilde Q_{1,2}}$.  \par
(b) There was no need to produce a large $A_t$ term in
Ref. \cite{split families unified}.  Due to the comparability of gauge
couplings in $G_1$ and $G_2$, and the low scale of $M_\Phi\sim 10^4$ GeV,
non-decoupling D term contribution to the Higgs mass could be significant.\par
(c) There, unification of gauge couplings was ``weak" in the sense of in
comparison with strong unification.

Here comes our final remarks.  First, it is worth pointing out that
despite the term strong ``unification", $SU(3)\times SU(2)\times U(1)$
do not necessarily unify into a larger simple group, so that there can
be no proton decay at all, and thus there is no the so-called
doublet-triplet splitting problem.  Because $g_2 \ll g_1$, the $G_2$
unification scale is the same as that of the traditional GUT, namely
about $3\times 10^{16}$ GeV.  Second, gauge couplings in $G_1$ are
expected to be a realization of GUT.  We have not studied that much
because it does not affect our physical results on one hand and the
couplings are too strong to use perturbation method on the other hand.
Third, LHC has set a lower bound on the gluino mass $m_{\tilde g}>1$ TeV
\cite{supersymmetry search}, and this bound would also apply to
$m_{\tilde Q_3}$ in traditional GMSB.  This is not the case for this
model, because $G_1\times G_2$ breaking also contributes the gluino an
additional mass $\sim \displaystyle \frac{g_2^2 V^2}{M}$ from mixing
with the fermionic component of $\Phi$.  This contribution is expected to
be larger than the purely soft mass
$\displaystyle\frac{g_2^2}{16\pi^2}\frac{F}{M}$ \cite{liu}.  Besides,
Higgs-mediated SUSY breaking contribution reduces
$m_{\tilde Q_3}$.  In a word, this model allows an interesting mass
pattern $m_{\tilde Q_{1,2}}\gg m_{\tilde g}\gg m_{\tilde Q_3}$.
Finally, in this work we have taken that the first two generations in
the same gauge group.  It is imaginable that we can introduce one more
version of the SM group to split these two generations further.  Namely
we may expect a model of $[SU(3)\times SU(2)\times U(1)]^3$, which first
breaks into $G_1\times G_2$ at some higher energy scale.

\begin{acknowledgments}

We would like to thank Jia-shu Lu for helpful discussions.
This work was supported in part by the National Natural Science
Foundation of China under Nos. 11375248 and 10821504, and by the
National Basic Research Program of China under Grant No. 2010CB833000.

\end{acknowledgments}

\end{document}